\documentstyle[psfig]{caosp}
\begin{document}
\pubyear{1998}
\volume{27}
\firstpage{374}
\htitle{Radiative transfer in Doppler Imaging}
\hauthor{N. Piskunov}
\title{Radiative transfer in Doppler Imaging}
\author{N. Piskunov}
\institute{Uppsala Astronomical Observatory, Uppsala, Sweden}

\maketitle

\begin{abstract}
The modern Doppler Imaging (DI) technique allows the reconstruction of
different stellar surface structures based on accurate calculation of
spectra of specific intensity. New applications like the mapping of the
magnetic field vector put very stringent requirements on the radiative
transfer (RT) solver which should be accurate, fast, and robust against
numerical errors. We describe the evaluation of three different algorithms
for our new magnetic DI code INVERS10. We also show the first results of
numerical experiments made with the new code. 

\end{abstract}

\section{Introduction}

{\em Doppler Imaging} is a method that allows to reconstruct the 
stellar surface structures from the rotational modulation of spectral 
line profiles (e.g. Piskunov \&\ Rice 1993). DI is an {\em iterative 
process}. On each iteration the observational data (spectra taken at 
different rotational phases) are computed for the current surface 
distribution at each observed wavelength $\lambda$ and rotational phase 
$\phi$. This is done by integrating the specific intensity $I_\lambda(M)$ 
over the visible hemisphere. {\bf $I_\lambda(M)$ must be computed at each 
iteration for each $\lambda$, $\phi$, and surface element $M$ in a
spectral range large enough to accommodate rotational Doppler shifts}. 
Therefore, spectral synthesis of specific intensities is the most time 
consuming part of any DI algorithm!

With the improvement of computer performance and the expansion of the DI
applications the need for a new RT became obvious as other solutions (e.g.
tables of pre-computed local profiles) have been exhausted. 

\section{Magnetic RT solver}

The goal of a magnetic DI code is to image the abundance (or temperature in
case of late-type stars) together with the magnetic vector -- 4 maps in
total to be reconstructed simultaneously. 

Three algorithms for RT solver in the presence of magnetic field have 
been considered: Runge-Kutta, Feautrier and Diagonal Element Lambda 
Operator (DELO). All three methods have been suggested and implemented by 
several people and detailed descriptions can be found in Landi 
Degl'Innocenti (1976, Runge-Kutta), Auer et al. (1977, 
Feautrier), and Rees et al. (1989, DELO). We strongly recommend 
those papers to anybody who wants to find out specific details about or 
implement any of these algorithms. Below we give a short outline of the 
magnetic RT problem and we describe the relevant properties of each 
algorithm.

The magnetic RT problem is a first order system of ordinary differential
equations:
\begin{equation}
\label{eq:RT}
\mu \frac {d{\bf I}} {d z} = -{\em K}{\bf I} + {\bf j},
\end{equation}
where ${\bf I} = (I, Q, U, V)$ is the vector of Stokes parameters, $\mu$
is the limb angle, ${\bf K}$ is the total absorption matrix and ${\bf j}$
is the total emission vector:
\begin{eqnarray}
{\em K} & = & \left(
\begin{array}{rrrr}
 {\em k}_c+{\em k}_l\cdot \phi_I & {\em k}_l\cdot \phi_Q &
           {\em k}_l\cdot \phi_U & {\em k}_l\cdot \phi_V \\

           {\em k}_l\cdot \phi_Q & {\em k}_c+{\em k}_l\cdot \phi_I &
           {\em k}_l\cdot \psi_V &-{\em k}_l\cdot \psi_U \\

           {\em k}_l\cdot \phi_U &-{\em k}_l\cdot \psi_V &
 {\em k}_c+{\em k}_l\cdot \phi_I & {\em k}_l\cdot \psi_Q \\

           {\em k}_l\cdot \phi_V & {\em k}_l\cdot \psi_U &           
-{\em k}_l\cdot \psi_Q & {\em k}_c+{\em k}_l\cdot \phi_I
\end{array} \right) \\
\nonumber \\
{\bf j} & = & \left( 
\begin{array}{r}
      {\em k}_c\cdot S_c + {\em k}_l\cdot S_l \phi_I \\
                           {\em k}_l\cdot S_l \phi_Q \\
                           {\em k}_l\cdot S_l \phi_U \\
                           {\em k}_l\cdot S_l \phi_V \end{array} \right), 
\end{eqnarray}
where ${\em k}_c$ and ${\em k}_l$ are the continuum and line opacity and 
$S_c$ and $S_l$ are the continuum and line source functions. Assuming no 
polarization in the continuum and LTE at the continuum formation depth,
$S_c$ is equal to the Planck function $B_\nu$. We note for later use that 
the diagonal elements of the absorption matrix are dominant, which 
provides the basis for the DELO algorithm.

The Zeeman splitting depends on the strength of the magnetic field and the
Land\'e factors of $\pi$- and $\sigma$-components. The amplitude of the
Stokes parameters depends on the orientation angles of the magnetic vector
(the angle $\gamma$ between the magnetic vector and the line of sight, and the
position angle $\chi$) via the absorption coefficients $\phi$'s and
anomalous dispersion coefficients $\psi$'s.  $\psi$'s are responsible for
magneto-optical effects. The relation of $\phi$'s and $\psi$'s to the line
profiles of the Zeeman components is given by:

\begin{eqnarray}
\phi_I & = & \frac{1}{2} \phi_p \sin^2\gamma + \frac{1}{4}(\phi_r+\phi_b)
             (1 + \cos^2\gamma) \nonumber \\
\phi_Q & = & \frac{1}{2}[\phi_p-\frac{1}{2}(\phi_r + \phi_b)]
             \sin^2\gamma\cos 2\chi \nonumber \\
\phi_U & = & \frac{1}{2}[\phi_p-\frac{1}{2}(\phi_r + \phi_b)]
             \sin^2\gamma\sin 2\chi \nonumber \\
\phi_V & = & \frac{1}{2}(\phi_r - \phi_b)] \cos\gamma \\
\psi_Q & = & \frac{1}{2}[\psi_p-\frac{1}{2}(\psi_r + \psi_b)]
             \sin^2\gamma\cos 2\chi \nonumber \\
\psi_U & = & \frac{1}{2}[\psi_p-\frac{1}{2}(\psi_r + \psi_b)]
             \sin^2\gamma\sin 2\chi \nonumber \\
\psi_V & = & \frac{1}{2}(\psi_r - \psi_b)] \cos\gamma \nonumber
\end{eqnarray}
where indices {\scriptsize p, b, r} stand for $\pi$-components and
{\em blue} and {\em red} $\sigma$-components.

The wavelength dependence of $\phi_p$, $\phi_b$, and $\phi_r$ are given 
by the Voigt function $V(a,v)$ while $\psi_p$, $\psi_b$, and $\psi_r$ are 
proportional to the Faraday-Voigt function $F(a,v)$. Huml\'{\i}\v{c}ek 
(1982) gives very fast and accurate complex approximation for 
$V(a,v)$ and $F(a,v)$. We have implemented it as FORTRAN and C routines
and compared it to a number of other approximations. We found
Huml\'{\i}\v{c}ek's approximation to be the best.

\subsection{Runge-Kutta magnetic RT integrator}

Runge-Kutta techniques for solving the radiative transfer equations 
(\ref{eq:RT}) integrate the Stokes parameters from the bottom of the 
atmosphere where an initial condition is set. A detailed description of 
the algorithm and its computer implementation (the MALIP code) has been 
given by Landi Degl'Innocenti (1976). He also analyses the main 
problems of the techniques. The advantage of Runge-Kutta is that the 
accuracy of the integration is checked at every step, so one can set the 
required accuracy {\em a priori}. We would also like to point out that 
the RT equation is one of a few rare cases where the 6th order 
Runge-Kutta offers substantial advantage over the conventional 4th order 
scheme because the accuracy can be checked without refining the step 
size. The main disadvantage is that different parts of the right hand 
side have a different depth dependence, and in order to achieve high 
accuracy, the algorithm is forced to use very small steps even deep in the 
atmosphere. To summarize: {\em the Runge-Kutta technique is accurate but 
slow. It is primarily useful as a reference for other methods.}

Now we shall turn to finite differences integration techniques which are 
more promising in terms of speed.

\subsection{Feautrier magnetic RT integrator}

The Feautrier method for solving the RT equation operates by splitting the 
intensity into two beams directed oppositely. The resulting equation is a 
second order ODE with two boundary conditions (one at the bottom and one 
at the surface of the atmosphere). Since the finite difference 
approximation involves 3 adjacent points for each step the method has 
excellent convergence properties. Application of the Feautrier 
method to non-magnetic RT requires the solution of a system of linear 
equations that form a tri-diagonal matrix. Although the accuracy cannot 
be checked at each step and the properties of the residual errors are 
much more complex than in the case of Runge-Kutta, refining the depth 
grid generally leads to a fast convergence and an accurate result. The 
Feautrier method has been extended to handle magnetic RT by Auer et al. 
(1977). In that case the tri-diagonal matrix is replaced by 
a block tri-diagonal, where each block is a 4$\times$4 matrix. The 
equations can be solved by analogy with the non-magnetic case, but 
back-substitution requires lots of 4$\times$4 matrix inversions and 
multiplications. The net result is a significant accumulation of 
numerical errors. For the centers of Zeeman components where the Voigt 
function is maximal and the Faraday-Voigt function is close to zero the 
difference between diagonal and non-diagonal elements in the absorption 
matrix reaches several orders of magnitude and with all the inversions, 
multiplications, and subtractions this scheme is bound to be numerically 
unstable. The alternative is to treat the block tri-diagonal matrix as a 
band diagonal matrix. The band should include 15 diagonals in order to 
cover all the blocks. The resulting scheme is robust against numerical 
errors for the price of only 20\,\% degradation in speed. Comparison with 
Runge-Kutta shows that for the same conditions, the Feautrier RT solver 
is about 30 times faster if the required accuracy is 10$^{-3}$.  That is 
not quite fast enough for MDI, as the typical disk integration procedure 
requires approximately 10$^3$ surface elements and the magnetic RT 
equation must be solved for each of them at several rotational phases.

\subsection{DELO magnetic RT solver}

Twelve years after the formulation of magnetic Feautrier algorithm, Rees 
et al. (1989) proposed a lambda operator methods serving as a 
one--way magnetic RT integrator. It is based on the fact that the 
absorption matrix is dominated by its diagonal elements. The principle 
can be easily illustrated in the non-magnetic case, but the DELO method 
is most impressive when integrating Stokes parameters.

In the non-magnetic case we can write the formal solution of the RT equation
connecting the intensities at optical depths $\tau_{k}$ and
$\tau_{k+1}$:
\begin{equation}
\label{eq:DELO}
I(\tau_{k})\ =\ I(\tau_{k+1})\cdot
    e^{\displaystyle (\tau_{k}-\tau_{k+1})}\ +\ \int_{\tau_{k}}^{\tau_{k+1}}
    e^{\displaystyle -(\tau-\tau_{k})}S(\tau)d\tau
\end{equation}
where $S(\tau)$ is the source function. If we assume that the source
function in our depth interval is linear in $\tau$ and can be expressed as
$S(\tau)=[(\tau_{k+1}-\tau)S_{k}+(\tau-\tau_{k})S_{k+1}]/
 (\tau_{k+1}-\tau_{k})$, then the integration in equation (\ref{eq:DELO})
can be performed analytically and we obtain a recurrence relation of the
type:
\begin{equation}
I(\tau_k)\ = \ P_k\ +\ Q_k\cdot I(\tau_{k+1})
\end{equation}
with a boundary condition at the bottom of the atmosphere.

Generalization to the magnetic case is straightforward. After we 
implemented this method, we found it to be free of numerical 
instabilities and about 6 times faster than the Feautrier method (both 
are a direct result of much fewer matrix inversions). On the down side, 
we found that the convergence properties of the DELO method are not as 
good as for Feautrier (not surprising as the latter is a second order 
finite difference method), and it takes a much finer grid (4 -- 8 times 
smaller step-size) to reach an accuracy of about 10$^{-3}$, thereby 
compromising the integration speed. After extensive experiments, we 
noticed that an adaptive depth grid can remedy the problem. The 
convergence is determined by the validity of linear approximation to the 
source function as described earlier in this section. It is much more 
efficient to refine the grid in the places where the source function 
variation is far from linear rather then adding extra points everywhere 
in the grid. Once implemented, this techniques proved to a be a winner. 
It usually takes as little as 20\,\% of additional grid points to reach 
the accuracy of 10$^{-3}$. {\em DELO with adaptive refinement of the 
depth grid is the fastest technique with good numerical stability and 
convergence properties}.

\begin{figure}[hbt]
\psfig{figure=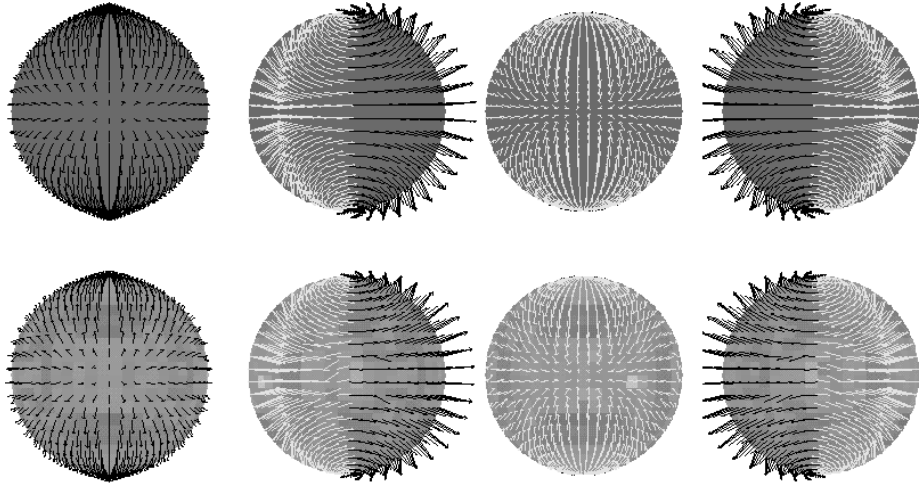,width=\textwidth}
\caption{The reconstruction of a magnetic dipole (bottom row) using 4 Stokes
         parameters. The original field structure is shown in the top
         row.}
\end{figure}

\section{The structure of INVERS10}

With the new powerful magnetic RT solver based on the DELO method, we are 
able to compute local Stokes profiles ``on the fly'' rather than 
pre-calculating the interpolation tables. For each rotational phase, our 
new MDI code computes the specific intensity (Stokes) profiles for the 
local magnetic field and local chemical composition, and derivatives with 
respect to field components (radial, and the two tangential) and the 
abundance: $\partial {\bf I}/\partial B_r$, $\partial {\bf I}/\partial 
B_m$, $\partial {\bf I}/\partial B_p$, and $\partial {\bf I}/\partial X$. 
The disk integration of the flux profiles takes into account the 
rotational Doppler shifts and the radial-tangential macroturbulence. 
After disk integration the discrepancy and the regularization functions 
are computed together with the gradient vector.  We use a modified 
conjugate gradient procedure to improve the solution. The modification 
makes use of the gradient vector during 1D optimization, since the 
gradient vector is computed with very little effort whenever the 
discrepancy function is evaluated. The overall procedure is efficient 
enough to reach a convergence for a typical size MDI problem (10 spectral 
lines, 100 wavelength points, 20 rotational phases) in 20-30 CPU hours on 
a fast workstation (an HP~9000 C-180 in our case) with about 15 minutes 
per function evaluation.

\begin{figure}[hbt]
\psfig{figure=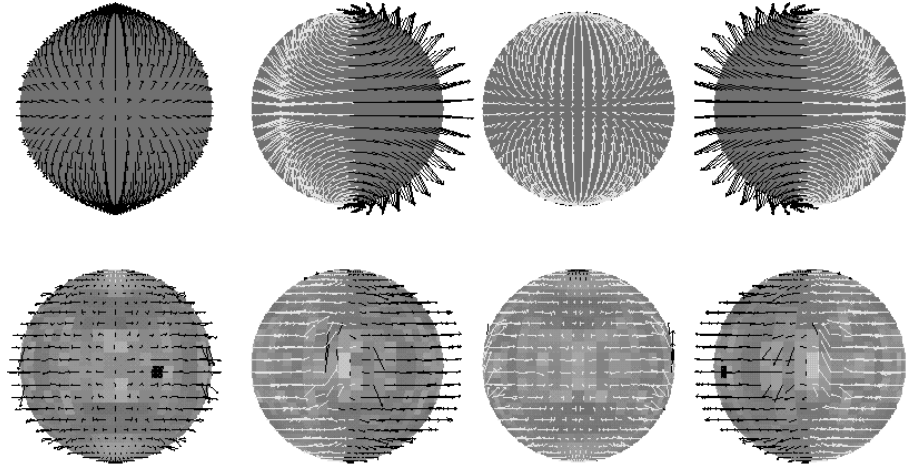,width=\textwidth}
\caption{The reconstruction of a magnetic dipole (bottom row) using only 2
         Stokes parameters ($I$ and $V$). The original field structure is
         shown in the top row.}
\end{figure}

\begin{figure}[hbt]
\psfig{figure=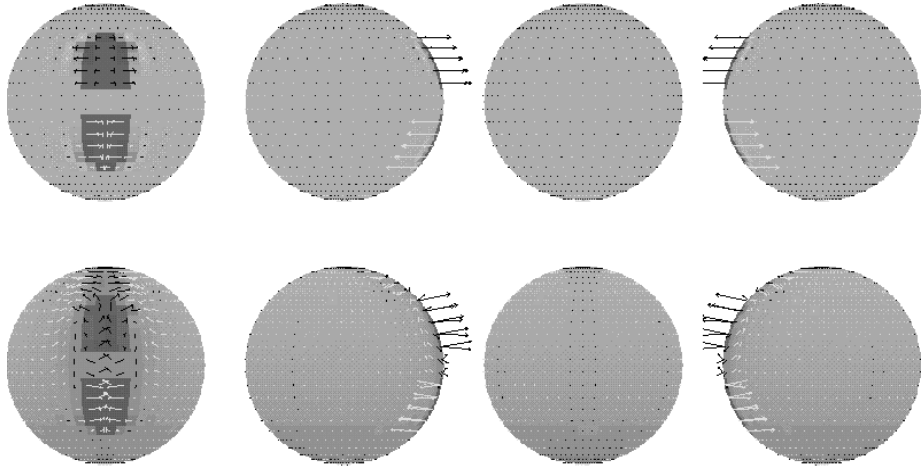,width=\textwidth}
\caption{The reconstruction of 2 magnetic spots with enhanced iron
         abundance (bottom row) using 4 Stokes parameters. The original
         field/abundance structure is shown in the top row.}
\end{figure}

\section{Numerical experiments}

Numerical experiments offer the best way to assess the reliability of an 
inverse code. We start with an artificial star with known surface 
structure, compute a set of ``observed'' profiles, and then use them as 
input data for the inversion. Below we show the results of 3 such 
experiments with INVERS10. In all cases we have used a rather typical 
Fe\,{\sc ii} 6141~\AA\ line, which has a Zeeman pattern with 6 $\pi$ and 
10 $\sigma$ components. The effective Land\'e factor is 1.5. The 
``observed'' profiles were computed for 20 equispaced rotational phases 
on a very fine surface grid using the Feautrier algorithm. The simulated 
profiles have been broadened by the instrumental profile corresponding to 
the resolving power of 80\,000 and mixed with random noise corresponding 
to S/N of 300 for $I$ and 1000 for $Q$, $U$, and $V$. The $v\sin i$ of 
the star was set to 30~km\,s$^{-1}$ with an inclination $i$ of 
45$^\circ$. Those parameters have been also used in the inversion. 
$\beta$ (in case of dipolar field) = 90$^\circ$, abundance contrast 
is 2\,dex.

In the first experiment (Fig. 1) we attempted to reconstruct the central
dipolar field. The magnetic axis was tilted by 90$^\circ$ from the
rotational axis and the polar field was 8000 Gauss. Chemical composition
was identical for every surface element. All four Stokes parameters were
used in the inversion. The initial guess had the correct chemical
composition but zero field. Figure~1 shows the results of successful
reconstruction.  The cross-talk between magnetic field and abundance map
of iron is less than 0.005\,dex in the abundance map and less then 200~G
in the magnetic map. 

In the next experiment (Fig. 2) we used the same test star, but only two
Stokes parameters ($I$ and $V$) were used in the inversion. The result is
shown in Figure~2. The reconstructed magnetic field differs significantly
from dipolar (most of the field vectors are directed along lines of
constant latitudes in stellar coordinates, lower panel on Fig.~2) while
the cross-talk reached the level of 0.5\,dex in the abundance map. 

In the last experiment (Fig. 3) two small spots of high (+2\,dex) iron
abundance were located at zero longitude with symmetrical placement
relative to the equator. Both spots have a radial magnetic field of 4000
Gauss, but opposite polarity. The results, shown in Figure~3, demonstrate
that 4 Stokes parameters, even with very modest phase coverage (as the
spots are visible only in 10 phases), can be used to recover realistic
field and chemical spot structures. 

\section{Conclusions}

Although many more experiments will be required to investigate all the 
properties of the new code, even now it is clear that we can {\em 
reliably reconstruct the vector magnetic field} and that {\em the 
observations of all four Stokes parameters are required}. It is also 
clear that the MDI problem must be solved in a consistent way rather then 
by separately imaging magnetic field and abundance (or temperature), 
since (at least with incomplete observations) one of the variables can 
successfully mimic the other.

\end{document}